\documentstyle[preprint,revtex]{aps}

\begin{document}
\thispagestyle{empty}


\begin{center}
\hfill{IP-ASTP-31-93}\\
\hfill{May, 1995}\\
\hfill{(Revised)}

\vspace{1 cm}

\begin{title}
Graviton Mode Function in Inflationary Cosmology
\end{title}
\vspace{1 cm}

\author{Kin-Wang Ng}
\vspace{0.5 cm}

\begin{instit}
Institute of Physics, Academia Sinica\\ Taipei, Taiwan 115, R.O.C.
\end{instit}
\end{center}
\vspace{0.5 cm}

\begin{abstract}
We consider the production of gravitons in an inflationary cosmology
by approximating each epoch of change in the equation of state as sudden,
from which a simple analytic graviton mode function has been derived. We use
this mode function to compute the graviton spectral energy density and the
tensor-induced cosmic microwave background anisotropy. The results are then
compared to the numerical calculations which incorporate a smooth
radiation-matter phase transition. We find that the sudden approximation is a
fairly good method. Besides, in determining the frequency range
and amplitude of the mode function, we introduce a pre-inflationary
radiation-dominated epoch and use a physically sensible regularization method.

\vspace{1 cm}
\end{abstract}
\newpage


\begin{center}
\bf{I. Introduction}
\end{center}

The spatial flatness and homogeneity of the present Universe strongly suggest
that a period of de Sitter expansion (inflation) had occurred in the
early Universe \cite{guth}. During inflation
quantum fluctuations of the inflaton field
may give rise to energy density (scalar) perturbations \cite{pi}, which can
serve as the seeds for structure formation of the Universe. Also,
a well-defined spectrum of gravitational waves could be
produced from the de Sitter vacuum \cite{star}. Since the waves decouple
from matter very early, they remain as a stochastic background of gravitational
waves at present and provide a potentially important probe of the early time.
Unfortunately, the detection of these primordial waves by using
terrestrial wave detectors or the timing of millisecond pulsars
\cite{krau} (this
is only sensitive to short-wavelength waves of period less than order years) is
still several orders of magnitude below the present experimental sensitivity.
However, being similar to energy density perturbations,
horizon-sized gravitational waves can induce
distortions of the cosmic microwave
background (CMB) \cite{ruba,star2,fabb,abbo,white,krau2,crit} via the
Sachs-Wolfe effect \cite{sach}.
In fact, large-angular-scale temperature anisotropy
of the CMB has been recently detected \cite{smoo}.
So, it is worthwhile to reexamine in detail the graviton production from
inflation.

Production of gravitational waves from quantum fluctuations during
inflationary period have been considered \cite{abbo,abbo2,alle,sah,mai}. In
transverse traceless (TT) gauge,
a classical gravitational wave has two independent polarization states. As
noticed by Grishchuk \cite{gris}, each polarization state of the wave behaves
as a
minimally coupled, massless, real scalar field, with a normalization factor of
$\sqrt{16\pi G}$ linking the two cases (we will show this in Section III).
Thus, the study of graviton production
during inflation reduces to considering the quantum fluctuations of a scalar
field in the de Sitter space-time. The result
is the well-known scale-invariant spectrum. The specrum can be derived by
relating
the scalar quantum-mechanical two-point function defined in the de Sitter
background to a two-point statistical average of an ensemble of
classical gravitational fields \cite{abbo}. In evaluating the two-point
function,
the de Sitter invariant vacuum has been chosen. The subsequent evolution of the
fields is then
governed by the classical equation of motion. The spectrum can also be derived
by a sequence of Bogoliubov transformations between creation and
annihilation operators defined in various stages of the Universe: inflationary,
radiation-, and matter-dominated. This consideration reveals that the results
are independent of state over a wide range of initial states in the
inflationary period, thus allowing one to select the de Sitter invariant
vacuum \cite{abbo2}.

In this paper, we will reexamine in detail the graviton production during
inflation.
We assume four sequential phases of the Universe:
radiation-, inflationary, radiation-, and matter-dominated. Each phase change
is approximated as a sudden process (i.e., assuming the transition from one
phase to another phase to be instantaneous).
We begin with a scalar field in a radiation-dominated period
prior to inflation, and work out the dynamics of the growing
quantum fluctuations of the field during inflation.
In evaluating the scalar two-point function, we avoid using any formal
renormalization schemes to remove divergences, instead, we will
adopt a physically sensible method introduced by Vilenkin to
regularize the infinities \cite{vile}. This allows us to determine
the frequency range and amplitude of the gravitons, and also
show explicitly the validity of selecting the de Sitter invariant vacuum during
the inflationary period. Based on the sudden approximation of the phase
changes, we will work out the graviton mode function for a wide
range of frequency which is relevant to astrophysical measurements.
This mode function is then applied to compute the spectral energy density and
the tensor-induced CMB anisotropy. Since in a realistic phase transition
the phase change is rather smooth than abrupt, we have to examine how good the
sudden approximation can reproduce the actual situation. In order to do so
we will compare the results obtained by taking
the sudden approximation to the numerical calculations which
incorporate a smooth phase transition.

\begin{center}
\bf{II. Background Metric of the Early Universe}
\end{center}

Let us suppose an initially radiation-dominated universe.
The Universe expanded and
became dominated by the energy density resided in the vacuum which drove
the inflation. After inflation, the Universe was reheated by the latent heat
released from the vacuum and resumed radiation-dominated. Later, matter
dominated the energy density and the Universe is currently in the
matter-dominated phase.
This supposition makes
sense only if inflation did occur below the Planck scale (for instance,
grand unified theories (GUT) inspired inflationary models), since a
temperature above
the Planck scale may not be defined. At temperatures well below the Planck
scale but above the GUT scale, radiation most likely
dominates the energy density of the Universe. For Planck-scale inflation (e.g.,
chaotic inflation \cite{lind}) or eternal inflation (inflationary period
extends to negative infinity),
it reduces to a three-step calculation in which the de Sitter invariant vacuum
in the inflationary epoch is reasonably singled out. In the following, we will
assume an inflation which is an exponential spatial expansion. In fact, our
results can be easily extended to other possibiliy like power-law inflation.
Since the curvature term is sub-dominated at very early times and the Universe
is essentially spatially flat after inflation, the space-time is well depicted
by the $k=0$ Robertson-Walker metric,

\begin{eqnarray}
ds^2 &=& dt^2-a^2(t) d{\bf x}^2 \nonumber \\
     &=& a^2(\eta) \left( d\eta^2-d{\bf x}^2 \right) \;,
\label{e1}
\end{eqnarray}
where $a(\eta)$ and $d\eta=dt/a(t)$ are the scale factor and conformal
time respectively. Here we are using the signature (-1,-1,-1,1).

Let us consider four phases: pre-inflation radiation-dominated,
inflationary, radiation-, and matter-dominated. The scale factor for each phase
is given by $a(\eta)=A(\eta+B)^{2/(1+3q)}$ according to the equation of state
$p=q\rho$ of that phase, where $A$ and $B$ are constants. For our
purposes, we normalize $a(\eta)=1$ and choose $\eta=-1/H$ at $t=0$ when
inflation begins, where
$H$ is the Hubble constant during inflation. Here we will
assume sudden phase transitions.
By requiring that $a(\eta)$ and ${\dot a}(\eta)$
(where dot means taking
derivative with respect to $\eta$) are continuous functions in the cosmological
history, we obtain

\begin{equation}
a(\eta)=\cases{2-\eta/\eta_3\;, & $\eta<\eta_3$\cr
               \eta_3/\eta\;,   & $\eta_3<\eta<\eta_2$\cr
       -\eta_3(\eta-2\eta_2)/\eta_2^2\;, & $\eta_2<\eta<\eta_1$\cr
                -\eta_3(\eta+\eta_1-4\eta_2)^2/4\eta_2^2(\eta_1-2\eta_2)\;, &
               $\eta_1<\eta$\cr}
\label{e2}
\end{equation}
where $\eta_3$, $\eta_2$,
and $\eta_1$ are respectively the conformal times at which the
inflationary, radiation-, and matter-dominated eras begin.
We have $\eta_3=-1/H$ and $\eta_2=-1/(H e^{H\tau})$ (where $\tau$ is the
duration of inflation). Note that $\eta\in (2\eta_3,+\infty)$
and today will be denoted by $\eta_0$. It can be easily
shown that

\begin{equation}
\eta_1\simeq -{{a(\eta_1)}\over{a(\eta_2)}}\eta_2\;,\;\;\;
\eta_0\simeq 2 \left[ {{a(\eta_0)}\over{a(\eta_1)}} \right]^{1\over 2}
\eta_1\;,
\label{e3}
\end{equation}
provided that $a(\eta_1)/a(\eta_2), a(\eta_0)/a(\eta_1) >>1$.

\begin{center}
\bf{III. Gravitational Wave Equation}
\end{center}

Weak gravitational waves on a Robertson-Walker universe has been investigated
by Lifshitz \cite{lifs}. As we restrict ourselves to a conformally flat
universe, this conformal property allows us to rederive the
wave equation in a much faster way as follows.

The theory of pure gravity is given by the Einstein-Hilbert action

\begin{equation}
I_G={1\over {16\pi G}} \int d^4 x \sqrt{g}\; R\;,
\label{e4}
\end{equation}
where $G$ is the gravitational constant.
In the weak field approximation, small metric perturbations are ripples on the
background metric:

\begin{equation}
g_{\mu\nu}=a^2(\eta)(\eta_{\mu\nu}+h_{\mu\nu})\;,\;\;\;h_{\mu\nu}<<1\;,
\label{e5}
\end{equation}
where $\eta_{\mu\nu}$ is the Minkowski metric, and Greek indices run from 0 to
3. In this section, we will leave $a(\eta)$ as a dimensionless and arbitrary
scale function. Therefore, without any loss of generality, we choose $a(\eta)$
very near to unity and write it as

\begin{equation}
a(\eta)=e^{\sigma(\eta)}\;;\;\;\;\sigma(\eta)<<1\;.
\label{e6}
\end{equation}
By expanding Eq.~(\ref{e4}) and keeping terms up to quadratic in
$h_{\mu\nu}$ and $\sigma$, we obtain

\begin{eqnarray}
I_G={1\over {16\pi G}} &\int& d^4 x\; a^2\; \Big ( {1\over 4} \partial_\mu
h_{\alpha\beta}
\partial^\mu h^{\alpha\beta} - {1\over 4}\partial_\mu h \partial^\mu h +
{1\over 2} \partial_\alpha
h^{\alpha\beta} \partial_\beta h \nonumber \\ &-& {1\over 2} \partial_\alpha
h^{\alpha\mu}
\partial^\beta h_{\beta\mu}    - 2\partial_\mu \sigma \partial^\mu h +
2\partial_\alpha
\sigma \partial_\beta h^{\alpha\beta} - 6\partial_\mu \sigma \partial^\mu
\sigma \Big ) \;, \label{e7}
\end{eqnarray}
where all indices are lowered and raised with $\eta_{\mu\nu}$ and
$h=\eta^{\mu\nu} h_{\mu\nu}$.
In synchronous gauge, $h_{00}=h_{0i}=0$, where $i$ runs from 1 to 3. The
remaining $h_{ij}$ contain a transverse, traceless tensor which corresponds to
a gravitational wave. Here we are only interested in this tensor mode.
Henceforth we will work in the TT gauge, i.e., $h^k_k=\partial_i h^{ij}=0$ and
denote the two independent polarization states of the wave as $+$, $\times$
\cite{misn}. Then, we obtain from Eq.~(\ref{e7}) the action of graviton as

\begin{equation}
I_{\rm graviton}={1\over {16\pi G}} \int d^4 x\; a^2\; {1\over 4}
\partial_\mu h_{ij} \partial^\mu h^{ij}\;.
\label{e8}
\end{equation}
We can write a monochromatic wave with a wave vector ${\bf k}$ as

\begin{equation}
h_{ij}(x)=h(x;{\bf k},\lambda)\;\epsilon_{ij}({\bf k};\lambda)\;,
\label{e9}
\end{equation}
where $\epsilon_{ij}({\bf k};\lambda)$ is the polarization tensor and
$\lambda=+,\times$. The polarization tensor satisfies

\begin{equation}
\epsilon_{ij}({\bf k};\lambda) \epsilon^{ij}({\bf k};\lambda')
=2\delta_{\lambda\lambda'}\;.
\label{e10}
\end{equation}
Hence, for this wave Eq.~(\ref{e8}) becomes

\begin{equation}
I_{\rm graviton}={1\over {16\pi G}}
\int d^4 x\; a^2(\eta)\; {1\over 2}
\left[
(\partial_\mu h(x;{\bf k},+))^2+ (\partial_\mu h(x;{\bf k},\times))^2
\right]\;,
\label{e11}
\end{equation}
which is in fact the action for two real, massless free scalar fields
$\phi(x;{\bf k},\lambda) = (16\pi G)^{-1/2} h(x;{\bf k},\lambda)$
in the background space-time. As a result, each
polarizaion state of the wave behaves as a real, massless, minimallly coupled
scalar field, with a normalization factor $\sqrt{16\pi G}$ relating the two
cases. Writing

\begin{equation}
h(x;{\bf k},\lambda)=(2\pi)^{-{3\over 2}}h_\lambda(\eta;{\bf k})\;e^{i{\bf
k\cdot x}}\;+\;{\rm h.c.}\;,
\label{e12}
\end{equation}
one finds from Eq.~(\ref{e11}) that the equation of motion for the wave
amplitude $h_\lambda(\eta;{\bf k})$ is

\begin{equation}
{\ddot {h_\lambda}}+2{\dot a \over a} {\dot {h_\lambda}}+k^2 h_\lambda =0\;,
\label{e13}
\end{equation}
where the dot denotes taking differentiation with respect to $\eta$. This is
just the Klein-Gordon equation for a massless plane wave in the background
space-time \cite{gris}. In what follows we will simply consider a massless real
scalar field $\phi(x)$.

\begin{center}
\bf{IV. Solutions of Wave Equation}
\end{center}

We consider the plane-wave mode expansion of a massless real scalar field in
the background metric~(\ref{e1}),

\begin{eqnarray}
\phi(x)&=&(2\pi)^{-{3\over 2}}\int d^3k \left[ a_{\bf k}\psi_k(\eta)\;e^{i
{\bf k\cdot x}}\;+\;{\rm h.c.}\right]\;, \nonumber \\
&&[a_{\bf k},a_{\bf k'}^\dagger] = \delta({\bf k}-{\bf k'})\;,
\label{e14}
\end{eqnarray}
where the mode function $\psi_k(\eta)$, which defines the vacuum state,
satisfies Eq.~(\ref{e13}).
When the scale factor $a(\eta)$ takes the form given by Eq.~(\ref{e2}), we
find:

\noindent{(i) When $\eta<\eta_3$,

\begin{equation}
\psi_k(\eta)=a^{-1}(\eta){1\over 2}(\pi|\xi|)^{1\over 2}
\left[ \alpha_1 H_{1\over 2}^{(1)}(k\xi) + \alpha_2 H_{1\over
2}^{(2)}(k\xi) \right]\;,
\label{e15}
\end{equation}
where $\xi=2/H+\eta$, $H_{1\over 2}^{(1)}$ and $H_{1\over 2}^{(2)}$ are
the Hankel functions.
The coefficients $\alpha_1$ and $\alpha_2$ are functions of $k$,
subject to the normalization condition:

\begin{equation}
|\alpha_2|^2 - |\alpha_1|^2 = 1\;.
\label{e16}
\end{equation}

\noindent{(ii) When $\eta_3<\eta<\eta_2$,

\begin{equation}
\psi_k(\eta)=a^{-1}(\eta){1\over 2}(\pi|\eta|)^{1\over 2}
\left[ \beta_1 H_{3\over 2}^{(1)}(k\eta) + \beta_2 H_{3\over
2}^{(2)}(k\eta) \right]\;.
\label{e17}
\end{equation}

\noindent{(iii) When $\eta_2<\eta<\eta_1$,

\begin{equation}
\psi_k(\eta)=a^{-1}(\eta){1\over 2}(\pi|\xi|)^{1\over 2}
\left[ \gamma_1 H_{1\over 2}^{(1)}(k\xi) + \gamma_2 H_{1\over
2}^{(2)}(k\xi) \right]\;,
\label{e18}
\end{equation}
where $\xi=\eta-2\eta_2$.

\noindent{(iv) When $\eta_1<\eta$,

\begin{equation}
\psi_k(\eta)=a^{-1}(\eta){1\over 2}(\pi|\xi|)^{1\over 2}
\left[ \delta_1 H_{3\over 2}^{(1)}(k\xi) + \delta_2 H_{3\over
2}^{(2)}(k\xi) \right]\;,
\label{e19}
\end{equation}
where $\xi=\eta+\eta_1-4\eta_2$. Similar to the case (i), the Greek
coefficients in cases (ii)-(iv) are $k$ dependent and each pair is subject
to the same normalization condition~(\ref{e16}). We list below the Hankel
functions
and their properties which will be useful for determining the coefficients:

\begin{eqnarray}
H_\nu^{(1,2)}(z)&=&J_\nu(z)\pm i N_\nu(z)=\sqrt{{2z}\over \pi}
h_{\nu-{1\over 2}}^{(1,2)}(z)
=\sqrt{{2z}\over \pi} \left[ j_{\nu-{1\over 2}}(z) \pm i n_{\nu-{1\over 2}}(z)
\right]\;, \nonumber \\
H_\nu^{(1)}(-z)&=&e^{-\nu\pi i} H_\nu^{(1)}(z)-2e^{-\nu\pi i}J_\nu(z)\;,
\nonumber \\
H_\nu^{(2)}(-z)&=&e^{-\nu\pi i} H_\nu^{(2)}(z)+2e^{\nu\pi i}J_\nu(z)\;,
\nonumber \\
H_{1\over 2}^{(1)}(z)&=&-i\sqrt{2\over{\pi z}} e^{iz}\;,\;\;
H_{1\over 2}^{(2)}(z)=i\sqrt{2\over{\pi z}} e^{-iz}\;, \nonumber \\
H_{3\over 2}^{(1)}(z)&=&-\sqrt{2\over{\pi z}} (1+{i\over z})\;e^{iz}\;,\;\;
H_{3\over 2}^{(2)}(z)=-\sqrt{2\over{\pi z}} (1-{i\over z})\;e^{-iz}\;,
\label{e20}
\end{eqnarray}
where $J_\nu$ and $N_\nu$ are Bessel functions, $j_{\nu-{1\over 2}}$ and
$n_{\nu-{1\over 2}}$ are spherical Bessel functions, $z>0$ is real, and $\nu$
is complex.

To determine the coefficients, we require that $\psi_k$ and its time
derivative $\dot \psi_k$ are continuous at $\eta$ equal to $\eta_3$, $\eta_2$,
and $\eta_1$. Thus, referring to Eqs.~(\ref{e15}-\ref{e19}), we have six
equations
with eight unknowns. However, for $\eta<\eta_3$, we choose the conformal vacuum
state, i.e., $\alpha_1=0$ and $\alpha_2=1$ in Eq.~(\ref{e15}),
which corresponds to the positive-frequency mode functions,
$\psi_k(\eta)=i\;a^{-1}(\eta)\;e^{-ik\xi}/\sqrt{2k}$, where
$\xi=2/H+\eta$. Note that these mode functions are conformal transforms of
the positive-frequency mode functions defined in Minkowski space-time.
As a result, the remaining six coefficients can be uniquely determined. In
terms of

\begin{equation}
z_3\equiv -k\eta_3={k\over H}\;,\;\;z_2\equiv -k\eta_2={k\over H}{1\over
a(\eta_2)}\;,\;\; z_1\equiv k(\eta_1-2\eta_2)={k\over H}
{{a(\eta_1)}\over{a^2(\eta_2)}}\;,
\label{e21}
\end{equation}
we find that

\begin{eqnarray}
\beta_1&=&-{1\over{2z_3^2}}\;, \nonumber \\
\beta_2&=&\left( 1-{i\over{z_3}}-{1\over{2z_3^2}} \right)\;e^{-2iz_3}\;,
\nonumber \\
\gamma_1&=&-\beta_1 \left( 1-{i\over{z_2}}-{1\over{2z_2^2}} \right)\;e^{-2iz_2}
- \beta_2 {1\over{2z_2^2}}\;, \nonumber \\
\gamma_2&=&\beta_1 {1\over{2z_2^2}} + \beta_2 \left(
1+{i\over{z_2}}-{1\over{2z_2^2}} \right)\;e^{2iz_2}\;, \nonumber \\
\delta_1&=&i\;e^{-iz_1}\gamma_1 \left( 1-{i\over{2z_1}}-{1\over{8z_1^2}}
\right) + i\;e^{-3iz_1}\gamma_2 {1\over{8z_1^2}}\;, \nonumber \\
\delta_2&=&-i\;e^{3iz_1}\gamma_1 {1\over{8z_1^2}} -
i\;e^{iz_1}\gamma_2\left( 1+{i\over{2z_1}}-{1\over{8z_1^2}} \right)\;.
\label{e22}
\end{eqnarray}

These results are obtained for any arbitrary $k$.
In the next section, by examining the dynamics of the growing quantum
fluctuations of the scalar field during inflationary period, we will show
that only plane-wave modes of $k$ within certain range will be excited and
generated from the de Sitter vacuum.

\begin{center}
\bf{V. Scalar Field Fluctuations}
\end{center}

The general
formalism for calculating the quantum fluctuation $\langle\phi^2\rangle$ in the
de Sitter space-time has been developed
by Bunch and Davies \cite{bun}. Using a point-splitting regularization scheme
they calculate a two-point function for $\phi(x)$ in Eq.~(\ref{e14}),

\begin{equation}
\langle\phi(x'')\phi(x')\rangle={1\over {(2\pi)^3}} \int_{-\infty}^\infty e^{i
\bf{k\cdot (x''-x')}} \psi_k(\eta'') \psi_k^*(\eta') d^3 k\;,
\label{e23}
\end{equation}
where $\psi_k(\eta)$, which defines the vacuum state, is given by
Eq.~(\ref{e17}).
Furthermore, they specialize to the de Sitter invariant vacuum such that
$\beta_1=0$ and $\beta_2=1$. With this choice, Eq.~(\ref{e23}) is singular for
$x''\rightarrow x'$. This singularity can be removed by an appropriate
renormalization scheme. In the present paper, we will treat
$\langle\phi^2\rangle$ based on an alternative procedure introduced by Vilenkin
\cite{vile}.

In an inflationary universe one has to carefully consider the physically
relevant quantity to be calculated. As has been discussed by Vilenkin, in the
inflationary universe the modes of interest are only those with wavelength
greater than the horizon, as these are those responsible for the growth of
$\langle\phi^2\rangle$. These are the modes with $k|\eta|<<1$, so that the
integration over
$k$ should be cut off at $k\simeq H e^{H\tau}$. When $k\le H$, the magnitude of
the fluctuation depends on the initial conditions of the universe, but with a
power-law expansion rate of the universe prior to inflation there is no reason
to believe that
$\langle\phi^2\rangle$ exceeds the anomalous de Sitter fluctuations
\cite{vile}. Hence we can write $\langle\phi^2\rangle$
at time $t$ as

\begin{equation}
\langle\phi^2\rangle=\langle\phi^2\rangle_0 + {1\over {2 \pi^2}} \int_H^{H
e^{Ht}} dk\;k^2 |\psi_k(\eta)|^2\;,
\label{e24}
\end{equation}
where $\langle\phi^2\rangle_0$ is the initial value and $t\le \tau$. Although
Eq.~(\ref{e24}) is not without
arbitrariness, Vilenkin's regularization tecnique makes physical sense as well
as being useful from a calculational point of view. In fact, this method can
easily reproduce previous results. For example, for a massless scalar
field, the growth of quantum fluctuations during inflation in this formalism,
by using Eqs.~(\ref{e17}), (\ref{e20})-(\ref{e22}), and (\ref{e24}), is
calculated as

\begin{eqnarray}
\langle\phi^2\rangle&=&{H^2\over{8\pi}}|\eta|^3 \int_H^{H e^{Ht}} dk\;k^2
|\beta_1 H_{3\over 2}^{(1)}(k\eta) + \beta_2 H_{3\over 2}^{(2)}(k\eta)|^2
\nonumber \\
&\rightarrow& {H^2\over{4\pi^2}}Ht\;\;{\rm as}\;\;t >> H^{-1}\;.
\label{e25}
\end{eqnarray}
This is the well-known result of the linear growth of fluctuations
\cite{pi,vile,vile2}. In essence, short-wavelength quantum
fluctuations in the scalar field $\psi_k$ of wavenumbers,
$H<k<He^{H\tau}$, will get red-shifted out of the horizon during the
inflationary period, after which they freeze in, remaining with constant
amplitude (see below). The superposition of these frozen modes therefore
constitutes the coherent scalar field. When the mode re-enters the horizon
much later
during the radiation- or matter-dominated
era it appears as a long-wavelength, classical wave.

\begin{center}
\bf{VI. Graviton Mode Function}
\end{center}

We write the graviton field in terms of the plane-wave modes given in
Eqs.~(\ref{e9}) and (\ref{e12}) as

\begin{eqnarray}
h_{ij}(x)&=&(2\pi)^{-{3\over 2}} \sum_\lambda \int d^3k
\left[ a_\lambda({\bf k})\;h_\lambda(\eta;k)\;e^{i{\bf k\cdot x}}\;
\epsilon_{ij}({\bf k};\lambda)\;+\;{\rm h.c.} \right]\;, \nonumber \\
&&[a_\lambda({\bf k}),a_{\lambda'}^\dagger({\bf k'})] = \delta({\bf k}-{\bf
k'})\; \delta_{\lambda\lambda'}\;.
\label{e26}
\end{eqnarray}
As explained above, the gravitational wave is related to the scalar field by
$h_\lambda(\eta;k)=(16\pi G)^{1/2} \psi_k(\eta)$. In the background metric~
(\ref{e1}),
$\psi_k(\eta)$ are given by Eqs.~(\ref{e15})-(\ref{e19}) with the coefficients
given by Eqs.~(\ref{e21})-(\ref{e22}). Similar to the scalar case, only quantum
fluctuations in the gravitational field $h_\lambda(\eta;k)$ of short
wavelengths, $H<k<H e^{H\tau}$,
will get pushed outside the horizon during the inflationary period. Therefore,
from Eq.~(\ref{e21}), $z_3$, $z_2$, and $z_1$ are such that

\begin{equation}
1<z_3<a(\eta_2)\;,\;\;\;{1\over{a(\eta_2)}}<z_2<1\;,\;\;\;
{{a(\eta_1)}\over{a^2(\eta_2)}}<z_1<{{a(\eta_1)}\over{a(\eta_2)}}\;.
\label{e27}
\end{equation}
In an inflationary cosmology, the exponential expansion factor $a(\eta_2)$ is
at least $10^{28}$ in order to circumvent the cosmological problems
\cite{guth}. Let
us assume $a(\eta_2)=10^{28}$. If the reheating temperature after inflation is
about the GUT scale of order $10^{15}{\rm GeV}$, then the ratio of the
scale factors at the beginning of matter-dominated era and at the end of
inflation, $a(\eta_1)/a(\eta_2)$, would be about $10^{23}$. In addition, the
present scale factor $a(\eta_0)\simeq 10^4 a(\eta_1)$ \cite{kolb}.
For definiteness, we fix

\begin{equation}
a(\eta_2)=10^{28}\;,\;\;\;{{a(\eta_1)}\over{a(\eta_2)}}=10^{23}\;,\;\;\;
{{a(\eta_0)}\over{a(\eta_1)}}=10^4\;.
\label{e28}
\end{equation}
Hence, we find from Eq.~(\ref{e27}) that

\begin{equation}
1<z_3<10^{28}\;,\;\;\;10^{-28}<z_2<1\;,\;\;\;10^{-5}<z_1<10^{23}\;.
\label{e29}
\end{equation}

It is useful to compare the size of a wave to the horizon size at time $\eta$.
For the background metric~(\ref{e1}), after inflation, the horizon size $l$
grows as \begin{equation}
l=a(\eta) \int_{\eta_2}^\eta d\eta \simeq a(\eta)\eta\;,\;\;\;\;
{\rm for}\;\;\eta>>\eta_2\;.
\label{e30}
\end{equation}
If we consider a wave with physical wavelength $\lambda_{\rm phys}$ which is
just entering the horizon at time $\eta$, then its wavenumber $k$ must satisfy

\begin{equation}
k={{2\pi\;a(\eta)}\over{\lambda_{\rm phys}}}={{2\pi\;a(\eta)}\over l}=
{{2\pi}\over \eta}\;.
\label{e31}
\end{equation}
This condition separates two limiting cases: a wave with $k\eta<<2\pi$ is well
outside the horizon whereas a wave with $k\eta>>2\pi$ is well within the
horizon.

Of particular interest are short-wavelength gravitational waves of period which
is within the
measurable range of terrestrial or astrophysical wave detector, as well as
waves
of long wavelengths comparable to the present horizon size. We can express the
wavenumbers $k$ of these waves in terms of the present time $\eta_0$ by
using Eq.~(\ref{e31}) as

\begin{equation}
k={{2n\pi}\over{\eta_0}}\;,
\label{e32}
\end{equation}
where $n=1$ corresponds to a wave just entering the present horizon, and the
wave with $n\ge 10^{10}$ (this wave re-enters the horizon during the
radiation-dominated epoch at time $\eta\le 10^{-10}\eta_0$) has
present period less than about a year. Substituting Eq.~(\ref{e32})
in Eq.~(\ref{e21}) and then using Eqs.~(\ref{e28}) and (\ref{e3}), we find for
such waves that

\begin{equation}
z_3\simeq 10^3 n\pi \;,\;\;\;z_2\simeq 10^{-25} n\pi \;,\;\;\;
z_1\simeq 10^{-2} n\pi \;.
\label{e33}
\end{equation}
Also, Eq.~(\ref{e29}) implies $10^{-3}<n<10^{25}$.
Therefore, in either case, $z_3 >> 1$ and $z_2 << 1$.
This leads to approximation of expressions~(\ref{e22}) as

\begin{eqnarray}
\beta_1 &\simeq& 0\;,\;\;\;\;\;\beta_2\simeq e^{-2iz_3}\;, \nonumber \\
\gamma_1 &\simeq& \gamma_2 \simeq -{{\beta_2}\over{2z_2^2}}\;,
\nonumber \\
\delta_1 &\simeq& \gamma_1 \left[ i\;e^{-iz_1} \left( 1-{i\over{2z_1}}-
{1\over{8z_1^2}} \right) + i\;e^{-3iz_1} {1\over{8z_1^2}} \right]\;,
\nonumber \\
\delta_2 &\simeq& \gamma_1 \left[ -i\;e^{iz_1} \left( 1+{i\over{2z_1}}-
{1\over{8z_1^2}} \right) - i\;e^{3iz_1} {1\over{8z_1^2}} \right]\;,
\label{e34}
\end{eqnarray}
where $z_3$, $z_2$ and $z_1$ are given in Eq.~(\ref{e33}). By using this and
Eqs.~(\ref{e17})-(\ref{e19}),
we can construct the wave form of the gravitational wave for any
particular value of $n$. For example, at time $\eta>>\eta_1$ in the
matter-dominated epoch, the wave amplitude is found to be given by

\begin{eqnarray}
h_\lambda(\eta;k) k^{3\over 2} \simeq &-&(32\pi G)^{1\over 2} H
{{k\eta_1}\over{k\eta}} e^{-2iz_3} \bigg\{
\biggl[ i\;e^{-iz_1} \biggl( 1-{i\over{2z_1}}-
{1\over{8z_1^2}} \biggr) + i\;e^{-3iz_1} {1\over{8z_1^2}}
\biggr] \nonumber \\  && h_1^{(1)}(k\eta)
+ \biggl[ -i\;e^{iz_1} \biggl( 1+{i\over{2z_1}}-
{1\over{8z_1^2}} \biggr) - i\;e^{3iz_1} {1\over{8z_1^2}} \biggr]
h_1^{(2)}(k\eta) \bigg\} \;,
\label{e35}
\end{eqnarray}
where we have used $\xi=\eta+\eta_1-4\eta_2 \simeq \eta$ in Eq.~(\ref{e19})
since $\eta>>\eta_1>>\eta_2$. In next section we will use this expression to
calculate the effects of primordial gravitational waves on the CMB. We note
that Eq.~(\ref{e35}) is at variance with previous results as found in Refs.~
\cite{ruba,star2,fabb,abbo,white}. However, we will show shortly
that those results can be in fact reproduced by taking different limits of
Eq.~(\ref{e35}). Also, similar expressions can be found in
Refs.~\cite{abbo2,alle,gris2,alle2}.
In the radiation-dominated epoch, for $\eta>>\eta_2$, we find that

\begin{equation}
h_\lambda(\eta;k) k^{3\over 2} \simeq -(8\pi G)^{1\over 2} H
e^{-2iz_3} j_0 (k\eta)\;,
\label{e36}
\end{equation}
which agrees with the result given in Refs. \cite{abbo2,white}.

As mentioned above, a gravitational wave well outside the horizon will remain
with constant amplitude until much later it re-enters the horizon as a
propagating classical wave. To illustrate this, we consider a
gravitational wave with wavelength larger than
the present horizon size ($n<1$).
For this wave, $z_1<10^{-2}\pi << 1$.
Thus, further approximation of Eq.~(\ref{e34}) can be made:

\begin{equation}
\delta_1 \simeq \delta_2 \simeq \gamma_1{3\over{4z_1}}\;,
\label{e37}
\end{equation}
where $\gamma_1$ is given in Eq.~(\ref{e34}). By using Eqs.~(\ref{e19})
and (\ref{e37}), we obtain the amplitude squared of the wave in the
matter-dominated epoch as

\begin{equation}
|h_\lambda(\xi;k)|^2 \simeq 16\pi G a^{-2}(\eta){{\pi\xi}\over 4}
4 |\delta_1|^2 {{2k\xi}\over \pi} j_1^2 (k\xi)\;,
\label{e38}
\end{equation}
where $\xi=\eta+\eta_1-4\eta_2$.
For $\eta>>\eta_1$, this can be simplified as

\begin{equation}
|h_\lambda(\eta;k)|^2 k^3 \simeq 8\pi G H^2
\left[ {{3j_1(k\eta)}\over{k\eta}} \right]^2\;,
\label{e39}
\end{equation}
where

\begin{equation}
{{3j_1(k\eta)}\over{k\eta}} \simeq 1\;\;\;{\rm as}\;\;\;n<<1\;.
\label{e40}
\end{equation}
The expression~(\ref{e39}) which is the long-wavelength limit of
Eqs.~(\ref{e19}) or (\ref{e35}) coincides with the well-known result of the
scale-invariant spectrum of gravitational waves generated by inflation
\cite{ruba,star2,fabb,abbo,white,abbo2}. After a wave has entered into the
horizon,
its wave amplitude decreases with time. To show this, we consider a wave with
$n=10^{10}$, which corresponds to $z_1=10^8\pi >> 1$.
Then, for $\eta>>\eta_1$, we find from Eq.~(\ref{e35}) that

\begin{equation}
|h_\lambda(\eta;k)|^2 k^3 \simeq 128\pi G H^2 (k\eta_1)^2
\left[ {{n_1(k\eta)}\over{k\eta}} \right]^2\;,
\label{e41}
\end{equation}
which agrees with the result given in Ref.~\cite{fabb} except differing by a
factor of 16. Note also that the $k$-dependence in Eq.~(\ref{e41})
differs from that in Eq.~(\ref{e39}), namely, the former has less power
suppression in $k$. This behavior has also been found by using transfer
function methods~\cite{turn}.

\begin{center}
\bf{VII. Sudden Transition Approximation}
\end{center}

The derivation of the graviton mode functions in Sec.~VI is based on the sudden
approximation of the phase changes. One should justify this approximation
before he can trust the result in Eq.~(\ref{e35}).
As seen in Eq.~(\ref{e2}), we have approximated each phase transition as being
instantaneous. In reality, changing from one phase to another phase must take
some finite time. The phase transition from the pre-inflationary
radiation-dominated era to the inflationary era and the reheating
at the end of inflation should involve very short
time scales, hence only affecting the extremely high frequency part of the
graviton spectrum.
Discussion of the influence of the reheating on the high frequency cutoff of
the spectrum can
be found in Ref.~\cite{moor}.
Here we will mainly concern the affect of the
radiation-matter phase transition on the comparatively much lower frequency
region of the spectrum. We will see that
this phase transition would affect the spectrum in the frequency range that
is relevant for CMB anisotropy calculation.

Many attempts have been made in the past in finding an accurate solution
for the graviton
mode function across the radiation-matter transition. In a two-component
(radiation plus dust) cosmology which has a smooth radiation-matter phase
transition, the equation of motion~(\ref{e13})
can be solved exactly in terms of complicated spheroidal wave functions
\cite{sah,kor}. In addition,
the equation can be solved approximately by the time-independent
\cite{turn} or time-dependent \cite{wang} transfer function method, or by the
WKB approximation \cite{ng}. Of course, the equation can be easily solved via
numerical method. However, sometimes it is much more convenient to have a
simple yet reasonably accurate graviton mode function. Undoubtedly, the sudden
approximation is the most economical way to obtain such a simple analytic
solution.

To justify the accuracy of the sudden approximation method,
we compare the period
of a wave, $2\pi/k$, with the characteristic duration of the radiation-matter
phase transiton, $\Delta\eta$, at time $\eta_1$. Unfortunately, the phase
transition,
being rather gradual than sudden, does not bear a characteristic timing.
However, $\Delta\eta$ should be comparable to $\eta_1$. As such,
the condition that the phase transition is sudden as seen by the wave is

\begin{equation}
{2\pi \over k} >> \Delta\eta\simeq \eta_1\;,\;\;\;{\rm
i.e.}\;,\;\;\;z_1<<2\pi\;.
\label{e42}
\end{equation}
Therefore, Eq.~(\ref{e35}) is valid for $z_1<<2\pi$, i.e., for waves entering
the horizon during the matter-dominated era. However, we should
look into this matter in more detail. Let us recast Eq.~(\ref{e13}) in terms of
the function $g_\lambda\equiv a(\eta) h_\lambda$ in the form

\begin{equation}
{\ddot {g_\lambda}}+\left(k^2 - {\ddot a \over a}\right) g_\lambda =0\;,
\label{e43}
\end{equation}
where

\begin{equation}
{\ddot a \over a}=\cases{0\;, & $\eta<\eta_1\;;$ \cr
                         2/\eta^2\;, & $\eta>\eta_1\;,$ \cr}
\label{e44}
\end{equation}
for a sudden radiation-matter phase transition.
In the radiation-dominated era, the solution to Eq.~(\ref{e43}) which matches
the wave function in the inflationary phase is given by $\sin(k\eta)$.
Hence, $h_\lambda k^{3\over 2} \propto \sin(k\eta)/(ka)$ (where $a\propto
\eta$), which is in fact the result in Eq.~(\ref{e36}).
In the matter-dominated era, the modes whose $k$'s satisfy
$k\eta\ge z_1>>{\sqrt 2}$ will not be affected by the term $\ddot a /a$.
Then, the solution to Eq.~(\ref{e43}) for these modes is again given by
$\sin(k\eta)$.
Hence, the matter-dominated mode function $h_\lambda$ is given by the same
solution in the radiation-dominated era with $a\propto \eta^2$ instead.
This actually reproduces the result in Eq.~(\ref{e41}).
Although the phase transition as
seen by the short-wavelength modes is gradual,
they are not affected by the sudden phase change at all and only scale as
$a(\eta)$. This is actually an illustration of the adiabatic theorem.
Therefore, the sudden approximation can give good approximate graviton mode
function across the radiation-matter phase transition for modes with
$z_1<<2\pi$ or $z_1>>{\sqrt 2}$. In the next section,
we will use Eq.~(\ref{e35})
to calculate the present spectral energy density of primordial gravitational
waves and the induced CMB anisotropy. These will be compared to the results
obtained by considering an equation of state which smoothly interpolates
between the radiation- and matter-dominated eras. We will find
that the sudden approximation can actually give fairly good approximation of
these quantities even in the regime $z_1\sim 1$.

\begin{center}
\bf{VIII. Spectral Energy Density and CMB Anisotropy}
\end{center}

To incorporate a smooth radiation-matter phase transition, we consider a
two-compontent universe containing radiation and dust. For convenience, we
define the new variables

\begin{equation}
\eta'\equiv (\sqrt 2 -1)\frac{\eta}{\eta_1}, \qquad k'\equiv
\frac{k\eta_1}{\sqrt 2 -1}.
\label{e45}
\end{equation}
It can be easily shown that

\begin{equation}
a(\eta')=a(\eta'_1)\eta'(\eta'+2).
\label{e46}
\end{equation}
Note that the radiation-matter equality time $\eta'_1\simeq 0.41$. For
$a(\eta'_0)/a(\eta'_1)=6000$,
the present time $\eta'_0\simeq 76.5$. (Note that $\eta'_0\simeq 63.5$ if we
have used Eq.~(\ref{e3}).) Then, the gravitational wave equation~(\ref{e13})
can be exactly solved in terms of complicated spheroidal wave
functions \cite{sah,kor}. Here we simply
solve it numerically, using the following initial conditions for $h_\lambda$:
\begin{equation}
h_\lambda k'\;^{3\over 2} =-e^{-2iz_3}(8\pi G)^{1\over 2}H \qquad{\rm
and}\qquad \frac{dh_\lambda}{d\eta'}=0\qquad{\rm as}\qquad
\eta'\to 0,
\label{e47}
\end{equation}
which is consistent with Eq.~(\ref{e36}).

{}From Eqs.~(\ref{e11}) and (\ref{e26}), the spectral energy density of
gravitational waves can be written as

\begin{equation}
\rho_g\equiv \sum_{\lambda=+,\times} k'\frac{d\rho_\lambda}{dk'}
=\sum_{\lambda=+,\times}
\frac{1}{32\pi G a^2(\eta')} \left(\frac{k'}{2\pi}\right)^3
\left[k'^2|h_\lambda|^2+\left\vert\frac
{d h_\lambda}{d \eta'}\right\vert^2 \right].
\label{e48}
\end{equation}
It is useful to define a dimensionless parameter which is the spectral energy
density divided by the closure density of the Universe,

\begin{equation}
\Omega_g\equiv \frac{\rho_g}{\rho_c},\qquad \rho_c=\frac{3H^2(\eta')}{8\pi G},
\label{e49}
\end{equation}
where the Hubble parameter $H(\eta')=(da/d\eta')/a^2$. As the graviton
production rate during inflation for each polarization state is statistically
equal, it is expected that $h_+ =h_\times \equiv -e^{-2iz_3}(8\pi G)^{1/2}H
k'^{-3/2}h$, where $h$ is real and normalized to unity as $\eta'\to 0$. Hence,

\begin{equation}
\Omega_g=\frac{v}{9\pi} \eta'^2 \left( \frac{\eta'+2}{\eta'+1}\right)^2
\left[k'^2 h^2+\left(\frac{dh}{d\eta'}\right)^2 \right],
\label{e50}
\end{equation}
where $v=3G H^2/8\pi=V_0/m_{Pl}^4$ is the inflation parameter
($V_0$ and $m_{Pl}$ are the
de Sitter vacuum energy and Planck mass respectively).
To estimate the amplitude of this quantity, one
typically makes use of the following physical argument. Before
a gravitational wave enters the horizon, its amplitude is constant.
After it enters the horizon, on the other hand, it behaves
effectively as radiation and will scale with $a$ as such. This gives
a $\Omega_g\simeq 10^{-13}$ for $v\simeq 10^{-9}$ \cite{alle,sah}.
Since we have already derived the graviton mode function, we can directly
calculate $\Omega_g$ at any time.
Fig.~1 shows the present time $\Omega_g$ in units of $v$ verus the
wavenumber in terms of $n$. Note again that $n=1$ corresponds to present
horizon-sized waves. The curves denoted respectively by the symbols $h_n$,
$h_s$, and $h_m$ are obtained by using the numerical solution of
Eq.~(\ref{e13}) with the initial conditions~(\ref{e47}), the sudden
approximation mode function~(\ref{e35}), and the matter-dominated mode
function~(\ref{e39}). We see that in the horizon- and super horizon-sized
wavelength region, the three curves are equal. While $h_m$ severely
underestimates the spectral energy density at large $n$'s, $h_s$ is a
fairly good approximation for all $n$'s. The worst underestimation by $h_s$
is by a factor of $2.5$ at $n\simeq 35$. At large $n$'s,
it approaches to a constant plateau and then falls off at the high frequency
cutoff \cite{kolb,moor}, being consistent with the numerical result.

In the last section, we have argued that
the sudden approximation should be valid for any $n$ which is much smaller or
larger than about $100$ (corresponding to $z_1\simeq \pi$). This is verified
in Fig.~1. It has been suggested that $\delta_1$ and $\delta_2$ in
Eq.~(\ref{e34}) for $z_1>2\pi$ (or $n>200$) should be modified to exponential
froms (see Eqs.~(4.22) and (4.23) of Ref.~\cite{alle2})
due to the adiabatic theorem.
However, we find that the spectral energy density obtained by using the sudden
approximation~(\ref{e35}) differs by less than $34\%$ for $n\ge 200$ from the
numerical result.

We now turn to calculate the CMB anisotropy induced by the primordial
gravitational wave background. The temperature anisotropy is induced via the
Sachs-Wolfe effect,

\begin{equation}
\frac{\delta T}{T}({\bf e}) =-{1\over 2}\int_{e}^{r}d\Lambda e^ie^j
\frac{\partial}{\partial\eta'}h_{ij}(\eta', \vec x),
\label{e51}
\end{equation}
where ${\bf e}$ is the propagation direction of the photon, $h_{ij}$ is
a field operator given by Eq.~(\ref{e26}), and the lower (upper) limit of
integration represents the point of emission (reception) of the photon.
In CMB measurements, the measured temperature anisotropy
is usually expanded in terms of spherical harmonics,

\begin{equation}
\frac{\delta T}{T}({\bf e}) = \sum_{l,m} a_{lm} Y_{lm}({\bf e}).
\label{e52}
\end{equation}
Since $\delta T/T$ arises from quantum fluctuations generated during inflation,
it is a Gaussian random field. This implies that $a_{lm}$'s are independent
Gaussian random variables satisfying

\begin{equation}
\langle a^{\dagger}_{lm} a_{l'm'}\rangle=\frac{C_l}{2l+1}
\delta_{ll'}\delta_{mm'},
\label{e53}
\end{equation}
where $C_l$ is the anisotropy power spectrum \cite{abbo}, from which we
can construct the two-point temperature correlation function,

\begin{equation}
\langle\frac{\delta T}{T}({\bf e}_1)\frac{\delta T}{T}({\bf e}_2)\rangle
 = \frac{1}{4\pi}\sum_l C_l P_l({\bf e}_1\cdot{\bf e}_2),
\label{e54}
\end{equation}
where $P_l$ is a Legendre polynomial. The formula for the power spectrum is
given by \cite{abbo}

\begin{equation}
C_l=\frac{32\pi v}{3} (2l+1)(l+2)(l+1)l(l-1)\int_0^\infty \frac{dk'}{k'}
    \left[\int_{\eta'_{dec}}^{\eta'_0} d\eta' \frac{dh(\eta')}{d\eta'}
    \frac{j_l[k'(\eta'_0-\eta')]}{k'^2(\eta'_0-\eta')^2}\right]^2,
\label{e55}
\end{equation}
where $j_l$ is a spherical Bessel function. We will use the decoupling time
$\eta'_{dec}\simeq 1.54$ corresponding to $a(\eta'_0)/a(\eta'_{dec})=1100$.
In Fig.~2, we plot $C_l$ for $l=2-250$ by using the three different
mode functions for $h$ as in the calculation of the spectral energy density.
Since the time integration in Eq.~(\ref{e55}) is getting close to the
radiation-matter equality time,
we should keep $\xi=\eta+\eta_1$ in Eqs.~(\ref{e35}) and (\ref{e39})
(i.e., replacing $\eta$ in these equations by $\eta+\eta_1$). We see from
Fig.~2 that the three curves almost match for $l<20$. The matter-dominated
mode function $h_m$ severely underestimates the $C_l$ for large $l$.
Using the sudden approximation mode function $h_s$, on the other hand, we
obtain much better accuracy; the $C_l$ we find differs by less than $2\%$
for $l\le 30$, less than $4\%$ for $l\le 130$, less than $10\%$ for $l\le 150$,
and less than $25\%$ for $l\le 250$ from the numerical result.

For a fixed $l$, the main contribution to the
integral~(\ref{e55}) for $C_l$ comes from the mode of wavenumber $k'\simeq
l/\eta'_0$ (or $l\simeq 2n\pi$) at the horizon crossing time $\eta'_c\simeq
\pi/k'$ \cite{star2,ng}.
In virtue of this, the mode functions $h_s$ and $h_m$, while underestimating
the spectral energy density at large $n$, underestimate the anisotropy power
spectrum at large $l$. It is interesting to note that in Fig.~2 all three
curves have minima at $l\simeq 170$. While both $h_n$ and $h_s$ have maxima at
$l\simeq 220$, $h_m$ has a maximum at a lower $l\simeq 210$.
It has been pointed out that it is the temporal phase of the
gravitational wave (the phase of $h$) which determines the overall shape of the
power spectrum at large $l$ \cite{ng}. Specifically, the location of a
maximum or a minimum is controlled by the phase of the wave. We thus plot
$h_n$,
$h_s$ and $h_m$ each of wavenumber $k'\simeq 2.2$ corresponding to $l\simeq
170$ at near the horizon crossing time $\eta'_c\simeq 1.4$ in Fig.~3. We find
that both $h_s$ and $h_m$, while underestimating the amplitude, reproduce
accurately the phase of the wave, in accordance with the behavior of the curves
at $l\simeq 170$ in Fig.~2. In Fig.~4, we plot again the three mode functions
with wavenumber $k'\simeq 2.9$ corresponding to $l\simeq 220$ at near $\eta'_c
\simeq 1.1$. We find that $h_s$ reproduces rather good the phase of the wave
and the location of the maximum at $l\simeq 220$. But, at
this $l$, not only $h_m$ badly underestimates both the amplitude of the wave
and the power spectrum, but also its $C_l$ curve rises up too quickly from the
minimum to the maximum.

The tensor-induced CMB anisotropy power spectrum has been calculated by using
numerical codes to evolve the photon distribution function using
the general relativistic Boltzmann equation for radiative transfer
\cite{crit,dod}, and by semi-analytic approaches using the Sachs-Wolfe
integral~(\ref{e55}). To evaluate the integral, different graviton mode
functions of varying accuracy have been used \cite{alle2,turn,kor,wang,ng}.
We find that the $h_n$ curve in Fig.~2 is very close to the result from
Boltzmann codes \cite{dod} as well as the result from spheroidal wave functions
\cite{kor}. Note that we have used $\eta'_{dec}\simeq 1.54$ in Eq.~(\ref{e55})
for all three cases. But, in the sudden approximation, Eq.~(\ref{e3}) implies
that $\eta'_{dec}$ should have been equal to $1.15$. However, in order to
make comparison between the three curves, we insist in using
a common $\eta'_{dec}\simeq 1.54$.
If we have used $\eta'_{dec}\simeq 1.15$ to evaluate the integral
$(\ref{e55})$,
the $h_s$ curve should have been similar to the results in \cite{alle2,kor}.
There they have also used a similar sudden approximation mode function to
calculate the power spectrum. Ref.~\cite{wang} has proposed an approximate
graviton mode function which in fact has an accuracy comparable to this work,
but has to involve more complicated time-dependent transfer functions.

\begin{center}
\bf{VIII. Discussions and Conclusions}
\end{center}

In this paper, we have attempted to derive a simple yet reasonably accurate
analytic mode function for the gravitational waves generated in
inflationary cosmology. The wave amplitude in Eq.~(\ref{e35}) is our main
result. To derive this result, firstly, we have assumed that
the Universe underwent
four phases in sequence: pre-inflation radiation-dominated, inflationary,
radiation-, and matter-dominated. This is quite natural as long as
one considers GUT-scale inflation. To the first order
approximation, we have treated each phase transition as being instantaneous.
Secondly, we follow Vilenkin's physically sensible method to
regularize the infinities in the evaluation of the scalar two-point function.
In essence, the regularization scheme simply introduces an infrared cutoff and
an ultraviolet cutoff. As a result, only those quantum fluctuations in the
gravitational field with wavenumbers lying between the two cutoffs
get pushed outside the horizon during the inflationary period,
after which they freeze out and become collective modes. Our four-phase
calculation explicitly shows that (see Eq.~(\ref{e22})) it is too naive
to select the de Sitter vacuum ($\beta_1=0,\beta_2=1$) during
the inflationary phase. Nevertheless, it could be a fairly good approximation
since $z_3>1$ (see Eq.~(\ref{e29})).
Moreover, for gravitational waves which are relevant to CMB observations,
pulsar timing measurements or terrestrial wave detectors, it is perfectly fine
to choose the de Sitter vacuum (see Eq.~(\ref{e34})). Besides, we can reproduce
previous results by taking certain limits of Eq.~(\ref{e35}). For
gravitational waves with large wavelength comparable to the present horizon
size, we have obtained Eq.~(\ref{e39}) which is the well-known scale-invariant
spectrum. For shorter waves, the wave form is different from
Eq.~(\ref{e39}), and only by choosing some special value of the
wavenumber can one obtain the result given in Eq.~(\ref{e41}).

We have used the sudden approximation mode function~(\ref{e35}) to calculate
the present graviton spectral energy density and the induced CMB anisotropy.
The results are compared to the results obtained by using the numerical
solution for the graviton mode function in a two-component universe which
incorporates a smooth radiation-matter phase transition. We have found that the
approximation of the phase change as a sudden process is a fairly good method.
Although the mode function~(\ref{e35}) mildly underestimates the spectral
energy density and the anisotropy power spectrum, it reproduces accurately the
overall shape of the power spectrum. This simple analytic mode function should
be quite useful in studying tensor perturbations from inflationary models.

\begin{center}
{\bf Acknowledgements}
\end{center}

This work was supported in part by the R.O.C. NSC Grant No.
NSC84-2112-M-001-024.

\newpage

\centerline{\bf Captions}
\vskip1.75truecm
\noindent{{\bf Figure 1}. Present graviton spectral energy density in units
of $v$. The symbols $h_n$, $h_s$, and $h_m$ denote respectively the curves
obtained by using the numerical mode function, the sudden approximation mode
function, and the matter-dominated mode function. Note that $n=1$ corresponds
to a wavelength of the present horizon size.
\vskip1.75truecm
\noindent{{\bf Figure 2}. Anisotropy power spectra by using $h_n$, $h_s$, and
$h_m$ respectively.
\vskip1.75truecm
\noindent{{\bf Figure 3}. Graviton mode functions each of wavenumber
$k'\simeq 2.2$ corresponding to $l\simeq 170$ at near the horizon crossing time
$\eta'_c\simeq 1.4$. The curves in descending amplitude are $h_n$, $h_s$, and
$h_m$ respectively.
\vskip1.75truecm
\noindent{{\bf Figure 4}. Graviton mode functions each of wavenumber
$k'\simeq 2.9$ corresponding to $l\simeq 220$ at near the horizon crossing time
$\eta'_c\simeq 1.1$. The curves in descending amplitude are $h_n$, $h_s$, and
$h_m$ respectively.}

\end{document}